\documentclass[structabstract]{aa}  
\usepackage{graphicx}
\usepackage{txfonts,natbib}
\usepackage{color}
\begin{document}
\title{Beryllium abundances along the evolutionary sequence of the open cluster 
IC~4651 -  A new test for hydrodynamical stellar models\thanks{Based 
on observations made with the ESO VLT, at Paranal Observatory, under programs 
065.L-0427 and 067.D-0126.}}

\author{R. Smiljanic \inst{1}\fnmsep\thanks{Current address: European Southern Observatory, Karl-Schwarzschild-Str. 2, 85748
        Garching bei M\"unchen, Germany. \email{rsmiljan@eso.org}}
        \and
        L. Pasquini\inst{2}
        \and
        C. Charbonnel\inst{3,4}
        \and
        N. Lagarde\inst{3}
}

\institute{Universidade de S\~ao Paulo, IAG, Dep. de Astronomia, Rua
     do Mat\~ao 1226, S\~ao Paulo-SP 05508-090, Brazil
\and
European Southern Observatory, Karl-Schwarzschild-Str. 2, 85748 Garching bei M\"unchen, Germany \\
\email{lpasquin@eso.org}
\and
Geneva Observatory, University of Geneva, chemin des 
Maillettes 51, CH-1290 Versoix, Switzerland \\
\email{Corinne.Charbonnel@unige.ch,Nadege.Lagarde@unige.ch}
\and
LATT, CNRS UMR 5572, Universit\'e de Toulouse, 
14 avenue Edouard Belin, F-31400 Toulouse Cedex 04, France
}

   \date{Received ; accepted }

\authorrunning{Smiljanic et al.}
\titlerunning{Beryllium abundances along the evolutionary sequence of IC 4651}

 
  \abstract
   {Previous analyses of lithium abundances in main sequence and red giant stars have 
revealed the action of mixing mechanisms other than convection in stellar interiors. 
Beryllium abundances in stars with Li abundance determinations can 
   offer valuable complementary information on the nature of these mechanisms. }
   {Our aim is to derive Be abundances along the whole evolutionary 
sequence of an open cluster. We focus on the well-studied open cluster IC~4651. These Be abundances 
are used with previously determined Li abundances, in the same sample stars, to investigate the 
mixing mechanisms in a range of stellar masses and evolutionary stages.}
   {Atmospheric parameters were adopted from a previous abundance analysis by the same authors. 
New Be abundances have been determined from high-resolution, high signal-to-noise UVES 
spectra using spectrum synthesis and model atmospheres. The careful synthetic modeling of the Be lines region
is used to calculate reliable abundances in rapidly rotating stars. The observed behavior of Be and Li is 
compared to theoretical predictions from stellar models including rotation-induced mixing, internal gravity waves, 
atomic diffusion, and thermohaline mixing.}
   {Beryllium is detected in all the main sequence and turn-off sample stars, both slow- and 
fast-rotating stars, including the Li-dip stars, but is not detected in the red giants. 
Confirming previous results, we find that the Li dip is also a Be dip, although the depletion of Be 
is more modest than for Li in the corresponding effective temperature range. 
For post-main-sequence stars, the Be dilution starts earlier within the Hertzsprung gap than expected from 
classical predictions, as does the Li dilution. A clear dispersion in the Be abundances is also observed. 
Theoretical stellar models including the hydrodynamical transport processes mentioned above 
are able to reproduce all the observed features well. These results show a good theoretical 
understanding of the Li and Be behavior along the color-magnitude diagram of this intermediate-age cluster 
for stars more massive than 1.2 M$_{\odot}$. }
   {}


   \maketitle
%

\section{Introduction}\label{sec:int}

The classical theory of stellar evolution, which only allows for mixing in 
stellar convective layers, fails to explain nearly all Li abundance 
patterns observed so far at the surface of low-mass stars. Indeed, in 
contradiction to the classical predictions, the vast majority of F- and 
early G-type stars (including the Sun) deplete, often quite severely, their 
surface Li abundance during their main-sequence lifetime, revealing
 transport processes beyond convection. A-type stars are also 
concerned, although in these objects the signatures of non-standard Li depletion, occurring 
in their interior during the main sequence, appear at their surface 
only when they cross the Hertzsprung gap and become subgiants.

Lithium depletion is seen both in field \citep[e.g.,][and references therein]{Herbig65,Duncan81,
PLP94,Leb99,Chen01,LRe04} and in open cluster stars \citep[e.g.][and references 
therein]{Wal65,Zappala72,Cayrel84,Soderblom90,
Balachandran95,Jones97,PRP01,SR05}. 
As cluster stars have well-defined masses and, most important, share the 
same age and initial chemical composition, they are ideal targets for investigating 
mixing processes as a function of stellar mass and evolutionary status.
They have therefore received considerable attention in the literature. 
Observations of the Li evolution along the color magnitude diagram (CMD) of 
open clusters have revealed a number of interesting features, such as the 
Li-dip \citep{Wal65,BT86a}, the lack of abundance trends 
with age among A-type stars \citep{BurCou98,BurCou00}, the presence of a 
spread of Li among solar-type stars in relatively old clusters like M67 \citep{PRP97,JFS99}, and the Li variations 
with mass and evolutionary status as in the case of IC 4651 \citep{PRZ04}.

Many different physical mechanisms have been proposed to explain these observations: 
atomic diffusion \citep{Mic86,Mi04}, mass loss \citep{Hobbs89,SwFa92}, 
rotation-induced mixing \citep{CVZ92,DP97,TC98,PTCF03,TV03},  
mixing by internal gravity waves \citep{GLS91,MoSc00,Young03}, or a 
combination of some of these processes. So far, only the 
hydrodynamic stellar models that combine the effects of meridional circulation, shear 
turbulence, internal gravity waves, and atomic diffusion have been able to 
account for Li observations over a broad range of stellar masses, ages, 
and evolutionary status \citep{TC03,TC05,ChT05,ipCT08}.
These models, which use a prescription for the wave excitation that reproduces 
the solar-p modes \citep{Goldreich94}, explain other observational 
constraints concomitantly, like the behavior of C, N, O, and Be across the Li dip in young 
clusters, such as the Hyades, or the internal rotation profile of the Sun \citep{ChT05}.

Additional clues can help ascertain the nature and description of transport 
mechanisms of chemicals and angular momentum inside low-mass stars. In particular 
the simultaneous determination of Li and Be abundances in the 
same stars strengthens the constraints by performing a stellar tomography. 
Such a powerful diagnosis is possible because Li and Be burn at different 
temperatures ($\sim$ 2.5$\times 10^{6}$ K for Li and $\sim$ 3.5$\times 10^{6}$ K for Be),
which correspond to different depths in the stellar interior \citep[see e.g.][]{ipDPC00}.

\begin{figure}
\centering
\includegraphics[width=7cm]{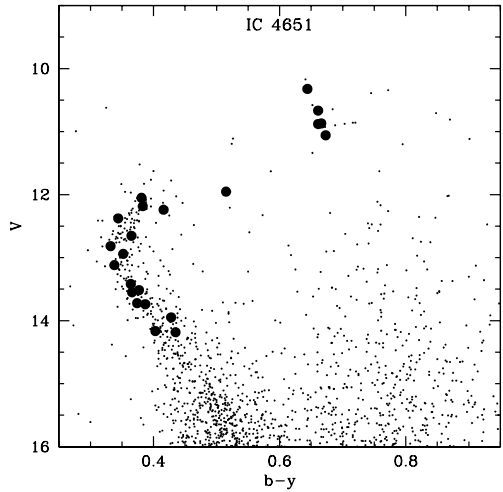}
\caption{Color magnitude diagram of IC 4651. Sample stars 
are shown as filled circles. The $ubvy$ photometry is from \citet{Mei00} 
and obtained by means of the WEBDA database.}
\label{fig:cmd}
\end{figure}

Although the simultaneous study of Li and Be in the same 
stars is a powerful tool, Be has been determined in far fewer stars 
than Li has. In particular, despite the advantages of cluster stars mentioned above, 
Be has been investigated in only a few open clusters \citep[e.g.][and references therein]{Boesgaard77,BAK04,
GL95,Ran02,Ran07}, while field stars have been more extensively studied \citep[e.g.][and references therein]{Boesgaard76b,SBKD97,San04b,BoKr09}.

Boesgaard and collaborators discovered a Be dip and a correlation between 
Li and Be depletion in mid F-stars in the field \citep{DBS98} and also in the Hyades, Coma, 
and Praesepe clusters \citep{BK02,BAK04}. They also show that for cooler G stars there 
is little or no Be depletion, even in Li-depleted objects up to $\sim$ 5500 K \citep[and references 
therein]{BAKDS04}. For even cooler stars, below $\sim$ 5500 K, \citet{San04b} 
show a Be decline correlated with decreasing temperature. These findings were confirmed by \citet{Ran02,Ran07}, who 
investigated Li and Be in late F- and early G-type stars in some open 
clusters over a relatively broad age range (IC 2391 -- 50 Myr, NGC 2516 -- 150 Myr, 
Hyades -- 600 Myr, IC 4651 -- 1.7 Gyr, and M67 -- 4.5 Gyr).

\begin{table}
\caption{Observational data of the sample stars.}
\label{tab:log}
\centering
\begin{tabular}{cccccc}
\noalign{\smallskip}
\hline\hline
\noalign{\smallskip}
Star & $v$  & ($b-y$) & obs. date & t$_{\rm exp}$ (s) & S/N \\
\hline
E3    & 12.05 & 0.38 & 10.may.2001 & 2 $\times$ 2100 & 90 \\
E5    & 12.82 & 0.33 & 09.may.2001 & 2 $\times$ 3000 & 90 \\
E7    & 14.17 & 0.40 & 26.mar.2000 & 4 $\times$ 3600 & 60 \\
E14   & 13.12 & 0.34 & 09.may.2001 & 2 $\times$ 3300 & 95 \\
E15   & 13.52 & 0.38 & 10.apr.2001 & 2 $\times$ 4500 & 85 \\
E19   & 12.24 & 0.42 & 08.apr.2001 & 2 $\times$ 2100 & 85 \\
E34   & 13.42 & 0.36 & 11.may.2001 & 2 $\times$ 4500 & 90 \\
E45   & 14.18 & 0.44 & 29.mar.2000 & 3 $\times$ 3600 & 40 \\
E56   & 12.18 & 0.38 & 08.jun.2001 & 2 $\times$ 2100 & 85 \\
E64   & 13.72 & 0.37 & 06.jun.2001 & 2 $\times$ 4500 & 100 \\
E79   & 13.55 & 0.37 & 09.apr.2001 & 2 $\times$ 4500 & 90 \\
E86   & 13.74 & 0.39 & 06.jun.2001 & 2 $\times$ 4500 & 100 \\
E95   & 11.95 & 0.52 & 08.jun.2001 & 2 $\times$ 2100 & 80 \\
E99   & 12.38 & 0.34 & 08.jun.2001 & 2 $\times$ 2100 & 90 \\
T1228 & 12.94 & 0.35 & 08.jun.2001 & 2 $\times$ 3300 & 90 \\
E25   & 12.65 & 0.37 & 07.jun.2001 & 2 $\times$ 2100 & 85 \\
T2105 & 13.95 & 0.43 & \multicolumn{3}{c}{\citet{Ran02}}  \\
\hline
E8    & 10.67 & 0.66 & 10.may.2001 & 1500 & 50 \\
E12   & 10.32 & 0.64 & 15.apr.2001 & 2 $\times$ 1200 & 75 \\
E60   & 10.87 & 0.67 & 09.apr.2001 & 2 $\times$ 1800 & 70 \\
E98   & 10.88 & 0.66 & 11.may.2001 & 2 $\times$ 1800 & 65 \\
T812  & 11.06 & 0.67 & 07.jun.2001 & 2 $\times$ 1800 & 70 \\
\noalign{\smallskip}
\hline
\multicolumn{6}{l}{The Str\"omgren photometry is from \citet{Mei00}. The S/N, per } \\
\multicolumn{6}{l}{resolution element (4 pixels), was measured in the Be region. The} \\
\multicolumn{6}{l}{ five stars grouped at the end of this and the following tables are the} \\
\multicolumn{6}{l}{ red giants.} \\
\end{tabular}
\end{table}

To improve our understanding of the mixing processes, adding new Be data covering a large fraction 
of the color magnitude diagram of well-studied open clusters is important. 
The ambitious aim is to compare the observations with models for stars over 
as wide as possible a range of mass and evolutionary status. 
In this context, the open cluster IC 4651 is a very good test case. 
It has well-determined membership \citep{Mei02}, and precisely 
known metallicity and chemical composition \citep[{[Fe/H]} = 
+0.11;][]{PRZ04,CBGT04,PP08}. Age determinations for this cluster vary 
between 1.2$\pm$0.2 Gyr \citep{Biaz07} and 1.7$\pm$0.15 Gyr \citep{Mei02}.
A color magnitude diagram of IC 4651 is shown in Fig. \ref{fig:cmd}. 

Important is that the Li evolution along the CMD of this cluster has already been 
investigated in detail by \citet{PRZ04}. Well-determined patterns 
have been found along its evolutionary sequence \citep[see Fig.~3 by][]{PRZ04}. 
In a sequence of increasing stellar mass, the cluster indeed has an Li ``plateau'' 
around solar mass stars, followed by a well-determined Li dip, after which the 
maximum Li content is observed, at meteoritic value. A sudden Li drop follows along the Hertzsprung gap, 
possibly indicating the onset of dilution (the so-called first dredge-up). 
Some of the giants show noticeable differences in Li abundance while sharing 
other stellar characteristics. It has been shown that models of rotating stars 
by \citet{CT99} and \citet{PTCF03} were able to reproduce the Li behavior 
both of main sequence stars lying on the blue side of the dip and of evolved stars.

\begin{table*}
\caption{Stellar parameters and abundances of Li and Be}
\label{tab:par}
\centering
\begin{tabular}{cccccccc}
\noalign{\smallskip}
\hline\hline
\noalign{\smallskip}
Star &  T$_{\rm{eff}}$ & Spec.   & log g & $\xi$ & $v~sini$ & A(Be) & A(Li) \\
     &   (K)         & or Phot. & (dex) & (km s$^{-1}$)  & (km s$^{-1}$) & (dex) &  (dex)    \\
\hline
E3  & 6550 & Phot. & 3.90 & 2.10 & 29.9 & 1.21  & 1.64 \\
E5  & 6930 & Phot. & 4.20 & 2.00 & 23.9 & 1.36  & 3.47 \\
E7  & 6300 & Spec. & 4.30 & 1.10 &  2.1 & 1.11  & 2.83 \\
E14 & 6860 & Phot. & 4.20 & 1.90 & 34.1 & 0.76  & $\leq$2.07 \\
E15 & 6540 & Phot. & 4.20 & 1.70 & 10.0 & 0.64  & $\leq$1.93 \\
E19 & 6280 & Phot. & 3.90 & 2.10 & 32.1 & 0.41  & $\leq$1.66 \\
E34 & 6640 & Phot. & 4.20 & 1.90 & 24.2 & 0.11  & $\leq$1.92 \\
E45 & 6350 & Spec. & 4.30 & 1.10 &  4.2 & 1.08  & 2.80 \\
E56 & 6520 & Phot. & 3.90 & 2.10 & 27.8 & 1.21  & $\leq$2.16 \\
E64 & 6650 & Spec. & 4.30 & 1.70 &  9.2 & 0.76  & 2.24  \\
E79 & 6620 & Spec. & 4.30 & 1.70 & 21.8 & 0.51  & $\leq$1.91 \\
E86 & 6600 & Spec. & 4.30 & 1.70 & 14.0 & 0.96  & 2.20  \\
E95 & 5800 & Spec. & 3.50 & 1.70 & 12.0 & 0.61  & 2.19  \\
E99 & 6830 & Phot. & 4.00 & 2.00 & 28.1 & 1.21  & $\leq$2.38 \\
T1228 & 6770 & Phot. & 4.20 & 1.70 & 5.7 & 1.21 & 3.18  \\
E25 & 6680 & Phot. & 4.00 & 2.00 & 21.3 & 1.11 &  3.29  \\
T2105 & 6110 & Phot. & 4.44 & 1.10 & 8.0 & 1.11 & 2.82 \\
\hline
E8  &  4900  & Spec. & 2.70 & 1.30 & 0.1 & -- & $\leq$0.09 \\
E12 &  5000  & Spec. & 2.70 & 1.50 & 1.1 & $\leq$$-$0.70 & $\leq$0.35 \\
E60 &  4900  & Spec. & 2.90 & 1.40 & 1.9 & $\leq$$-$0.70 & 0.42 \\
E98 &  4900  & Spec. & 3.00 & 1.40 & 1.0 & $\leq$$-$0.50 & 0.72 \\
T812 &  5000 & Spec. & 3.00 & 1.60 & 0.5 & $\leq$$-$0.70 & $\leq$0.38 \\
\noalign{\smallskip}
\hline
\end{tabular}
\end{table*}

In this work we present Be abundances for the same stars analyzed by
\citet{PRZ04}. This is the first time Be abundances 
have been determined for stars along the whole evolutionary sequence of 
an open cluster, including solar-type, Li-dip, turn-off, subgiant, and red giant stars. 
With these new results, the mixing processes 
in different stellar masses can be investigated in greater detail.

\section{Observations and stellar parameters}\label{sec:obs}

Our sample is composed of 22 stars. Twenty one of them have been analyzed 
by \citet{PRZ04}. To these we add one star of this same cluster 
analyzed by \citet{Ran02}, star T2105. The sample was selected to cover 
a large interval of the CMD from the Str\"omgren photometry obtained 
by \citet{Mei00}. The interested reader is referred to the work of 
\citet{PRZ04} for more details on the sample selection.  

Spectra were obtained using UVES, the \emph{Ultraviolet 
and Visual Echelle Spectrograph} \citep{ipDe00} fed 
by UT2 of the VLT. UVES is a cross-dispersed echelle spectrograph 
able to obtain spectra from the atmospheric cut-off at 300 nm to 
$\sim$ 1100 nm.  The reduction was conducted with the ESO UVES 
pipeline within MIDAS. The spectra have typical signal-to-noise (S/N) between 
40 and 100 and R $\sim$ 45\,000. The log book of the observations 
is given in Table \ref{tab:log}. Stars with names starting with an ``E'' 
follow the numbering system of \citet{Egg71}. Stars with names starting 
with a ``T'' follow the numbering system of \citet{AT88}. 

The atmospheric parameters for the sample stars determined by 
\citet{PRZ04} were adopted. A first estimate 
of the effective temperature (T$_{\rm eff}$) was calculated using 
the calibrations of \citet{AAM96b,AAM99}. A first estimate of the 
surface gravity was obtained using this T$_{\rm eff}$, a mass of 
1.80 M$_{\odot}$ for the cluster turn-off, and the distance modulus 
determined by \citet{Mei02}, (m-M)$_{0}$ = 10.03. Starting with these 
values, both the T$_{\rm eff}$ and the microturbulence velocity ($\xi$) were 
further constrained using \ion{Fe}{i} lines, eliminating trends with 
the excitation potential and the equivalent widths, respectively. 
The ionization equilibrium, on the other hand, was not adopted 
to constrain the gravity. While the other parameters were being 
varied, gravity was also varied by at most 0.30 dex to obtain 
as good an equilibrium as possible without departing much from 
the log g given by the star position at the CMD.

Using this method, \citet{PRZ04} had some difficulty 
constraining the parameters of the fast-rotating hot stars around the 
turn-off. The above-mentioned approach resulted in higher metallicities 
for these stars than the ones determined for the slow-rotating stars. This effect was likely 
caused by the blending of the \ion{Fe}{i} lines with other neighboring 
lines due to the rotational broadening. Thus, the measured 
equivalent widths were greater than they should be. As in these cases 
the spectroscopic parameters are likely not reliable, atmospheric 
parameters for the fast-rotating stars were calculated adopting 
the photometric temperatures. The spectroscopic temperatures were 
adopted for the remaining stars. The atmospheric 
parameters of the stars are given in Table \ref{tab:par}. 
The mean metallicity of the cluster as determined by \citet{PRZ04} 
was adopted for all the stars, [Fe/H] = +0.11.

\section{Be abundances}\label{sec:abun}

\subsection{Synthetic spectra}

For late-type stars, Be abundances can only be determined using  
the \ion{Be}{ii} $^{2}$S - $^{2}$P$_{0}$ resonance lines at 3131.065 \AA\@ 
and 3130.420 \AA. This near-UV region is extremely crowded with atomic 
and molecular lines, some of them still lacking proper identification. 
Thus, determination of Be abundances needs to be conducted with spectrum 
synthesis taking all the blending nearby features into account.

\begin{figure}
\begin{centering}
\includegraphics[width=7cm]{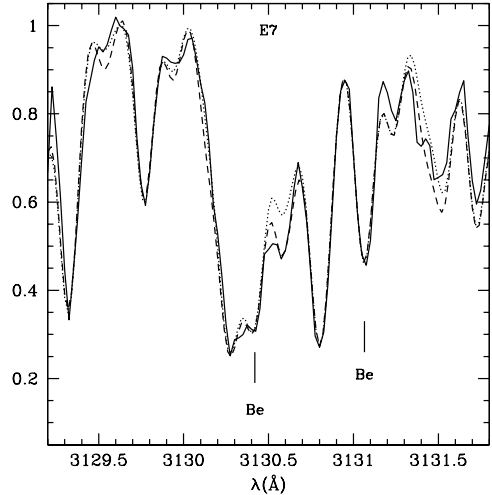}
\caption{Fit to the region of the Be lines in the main 
sequence star E7. The solid line represents the observed spectrum, the 
dotted line a synthetic one with solar abundance ratios, and the dashed line 
a synthetic spectrum with [O/Fe] = +0.25 and [C/Fe] = +0.10.} 
\label{fig:e7}
\end{centering}
\end{figure}

To derive the Be abundances, synthetic spectra were calculated with the
codes described in \citet{CB05} and the grids of
model atmospheres without overshooting calculated with the ATLAS9
program \citep{ipCK03}. These models assume local
thermodynamic equilibrium, plane-parallel geometry, and hydrostatic
equilibrium. 

The line list is the same as used in \citet{Sm08,Sm09}. 
In short, the molecular line list is described in \citet{CB05} and the atomic line
list is the one compiled by \citet{Pr97}.  Log $gf$ of $-$0.168 and
$-$0.468 were adopted for the Be lines at 3131.065 \AA\@ and 3130.420 \AA, respectively. The line list includes an
\ion{Fe}{i} line in 3131.043 \AA, with log $gf$ = $-$2.517 and $\chi$
= 2.85 eV, affecting the blue wing of the Be 3131 line. The parameters
of this line were constrained using several stars of different
parameters and metallicities \citep[see][for details]{Pr97}. The 
proper identification of this line, however, remains
controversial in the literature, and other choices have been made by
different authors. As discussed in more detail by \citet{Ran07}, 
this choice has no significant effect on the calculation of Be abundances of stars with
T$_{\rm eff}$ $>$ 5400 K. 

\subsection{The Sun and the slow-rotating stars}

That some of the sample stars are fast rotators adds further 
difficulty to the analysis of the Be spectral region. Thus, we conducted a 
careful analysis, first modeling the slow-rotating stars and conducting 
a few tests to make certain any possible systematic effect affecting 
the modeling of the fast-rotating stars would be under control.

As the first step, with the line list described above, we fitted the 
Be lines in the solar UVES\footnote{The spectrum is available for 
download at the ESO website:
  www.eso.org/observing/dfo/quality/UVES/pipeline/solar\_spectrum
  .html} spectrum, adopting the parameters: T$_{\rm eff}$ = 5777 K, log
g = 4.44, and $\xi$ = 1.00 km s$^{-1}$. An abundance of A(Be) = 1.10 was 
obtained. This abundance is in excellent agreement with the one 
found by \citet{Chm75}, A(Be) = 1.15, usually adopted as the reference
photospheric solar abundance.

The low photospheric solar Li abundance, depleted by a factor of $\sim$ 160 
with respect to the meteoritic value, has been known for a long time \citep[see e.g.][]{Mul75}. 
The photospheric Be abundance and its difference with respect to the meteoritic one have instead been  
a matter of debate. All analyses of the solar photospheric 
abundance obtain log(Be/H) $\sim$ 1.15 -- 1.25 \citep{Chm75,GL95,Ran02}, 
a factor of 2 less than what is measured in meteorites, log(Be/H) = 1.41 
\citep{Lod03}. This issue now seems settled to a large extent; the
measured photospheric value is lower than the real one due to our 
inability of properly accounting all continuum opacity sources in the region 
around 313 nm, where the two \ion{Be}{ii} resonance lines are located, when
calculating model atmospheres \citep{BaBe98,Asp04}.

The second step was to fit the two main sequence slow rotator stars E7 and
E45. The cluster was found to have a solar
abundance pattern by \citet{PRZ04}. Thus, the solar abundance 
ratios\footnote{The abundances recommended by \citet{GS98} were 
adopted for all elements except for oxygen, for which we adopted the
 abundance suitable for the 1D atmospheric models recommended by
 \citet{ALA01}, A(O) = 8.77, and for Be for which the abundance 
derived above from the solar spectrum was adopted.} 
were adopted when calculating a first set of synthetic spectra. 
However, as shown in Fig.\@ \ref{fig:e7}, when adopting a solar 
abundance pattern, an important molecular feature near the 
3130 \AA\@ Be line is not properly fitted. A better fit is obtained 
when [O/Fe] is increased by +0.25 dex and [C/Fe] by +0.10 dex, 
in both E7 and E45. These apparent overabundances are likely not 
to be real but caused by shortcomings in the analysis, such as the 
neglect of effects caused by departures from the local 
thermodynamic equilibrium and/or by problems with the molecular 
and atomic data in this region.  

Nevertheless, it is important to notice that properly fitting the 
neighboring molecular features in the slow-rotating stars is an important 
step toward analyzing the fast-rotating ones. In the last case, all these 
lines are blended into a single feature. As all the stars in the cluster are 
expected to have the same chemical composition (except for the giants), 
one may expect that the enhanced abundances of C and O are also 
needed to properly fit the spectra of the fast-rotating stars. We stress that 
this increase in the abundances of C and O has no effect on the calculated 
Be abundances of the slow-rotating stars. Thus, when fitting the fast rotators, C and O abundances were 
kept with these increased values, while the other elements were 
kept with a solar ratio. Only the Be abundance was varied from star to star. 
The derived Be abundances are listed in Table \ref{tab:par}. 

\subsection{The fast-rotating stars}

\begin{figure*}
\begin{centering}
\includegraphics[width=10cm]{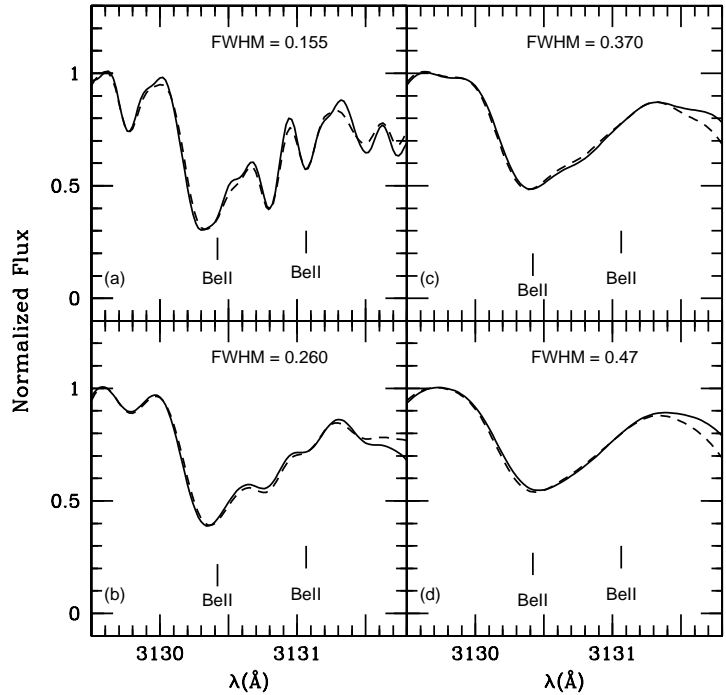}
\caption{Fit to the region of the Be lines in the main 
sequence star E7. The four panels show a sequence of larger broadening 
applied to the observed spectrum. In all panels the best-fitting synthetic 
spectrum has A(Be) = 1.16 dex.}
\label{fig:gauss}
\end{centering}
\end{figure*}

Before fitting the true fast-rotating stars, we conducted a test to 
better understand the effects of the broadening in the derived 
abundances. We artificially broadened the spectra of the two 
slow-rotating stars, E7 and E45, by convolving them with Gaussian 
functions. In Fig. \ref{fig:gauss} we show the spectrum of the star 
E7 broadened with four different Gaussian functions. The Be abundance 
was then calculated for these broadened spectra. 

In the four cases shown in Fig.\@ \ref{fig:gauss}, an abundance of
A(Be) = 1.16 dex was derived for E7, while A(Be) = 1.11 dex was
derived in the original unbroadened spectrum. Applying similar broadening
to E45, we obtain A(Be) = 1.15 for all the broadened cases, while A(Be)
= 1.08 dex was obtained for the original observed spectrum. Thus, we
conclude that the abundances in a broadened spectrum are slightly 
overestimated. However, this small amount is well within the
uncertainties (see below). Therefore, we are confident that it is
possible to derive meaningful abundances for
the fast-rotating stars with our method of analysis.  

\begin{figure}
\begin{centering}
\includegraphics[width=7cm]{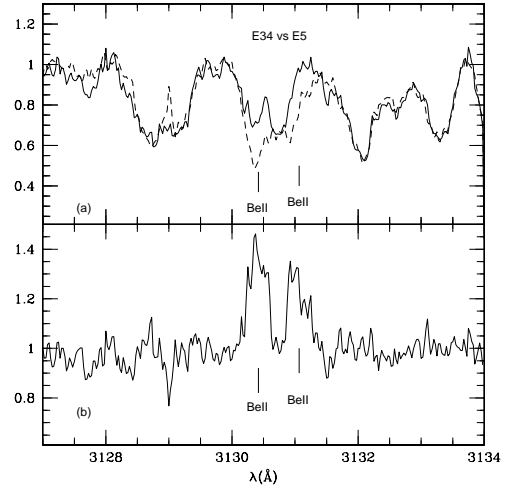}
\caption{(a) Comparison between the observed spectra of stars E5 (dashed line) and 
E34 (solid line). The stars have similar rotational velocity and different Be abundances. 
(b) The result of the division of the two spectra in the upper
panel. The two residual features are located at 3130.41 and 3131.03
\AA. These wavelengths are compatible with the difference being caused by 
different Be abundances.} 
\label{fig:div}
\end{centering}
\end{figure}

\begin{figure}
\begin{centering}
\includegraphics[width=7cm]{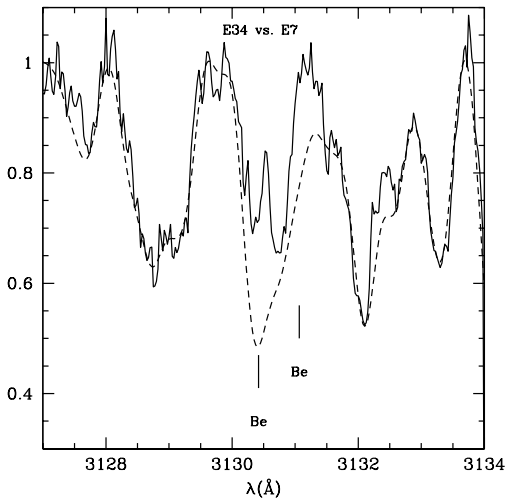}
\caption{Comparison between the observed spectrum of star E34 (solid line) and 
the spectrum of star E7 (dashed line) artificially broadened to a $v~sini$ similar to 
the one of E34. It is clear the lower Be abundance of star E34 and that a higher 
abundance could be easily detected in our analysis.}
\label{fig:e34e7}
\end{centering}
\end{figure}

Beryllium abundances were then derived for all the fast-rotating 
main-sequence and turn-off stars. The stars were found to have a broad 
range in abundances, from 1.36 dex (E5) to 0.11 dex (E34). In Fig.\@ 
\ref{fig:div} we plot the spectra of these two stars, which have 
similar $v~sini$, to show that the difference in Be abundance obtained 
in the analysis is real. In the lower panel of this same figure 
we show the result of the division of these spectra. The structures 
left in the division have central wavelengths of 3130.41 and 3131.03 \AA, 
arguing that the difference is really caused by the different Be abundances. Similarly, 
we compare in Fig.\@ \ref{fig:e34e7} the observed spectrum of star 
E34 and the spectrum of E7 artificially broadened to a $v~sini$ similar to that 
of E34. It is clear from this figure that star E34 has a smaller Be abundance 
than E7. This comparison again shows that the difference in Be abundances 
detected in the stars is real.


\begin{figure}
\begin{centering}
\includegraphics[width=7cm]{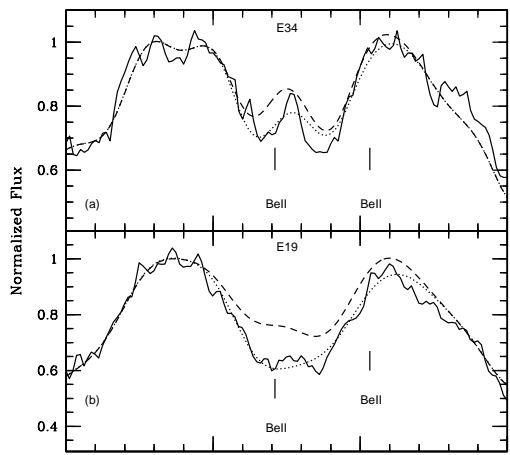}
\caption{(a) Comparison between the observed spectrum of star E34 (solid line), the 
best synthetic fit (dotted line), and a synthetic spectrum calculated without Be (dashed line). 
(b) Comparison between the observed spectrum of star E19, the 
best synthetic fit, and a synthetic spectrum calculated without Be.
These two comparisons argue for both stars having detectable amounts 
of Be.} 
\label{fig:e34e19}
\end{centering}
\end{figure}

Finally, we point out that Be was detected in all the sample of main 
sequence and turn-off stars. Most of the stars show signs of Be depletion, i.e., 
Be lower than the meteoritic value, but still have a detectable amount of Be in their photospheres. In Fig.\@ 
\ref{fig:e34e19} we show the observed spectra of E19 and E34, the two 
stars with smaller Be content. Also shown are a best-fit synthetic spectrum 
together with a spectrum calculated with no Be at all. As clearly seen, 
the spectra lacking Be show a marked difference when compared to the observed 
ones. This comparison argues that our Be abundances are actual detections 
and not upper-limits. The derived Be abundances are listed in Table \ref{tab:par}. 

\subsection{The red giants}

Fitting synthetic spectra in the Be region for the red giants proved to be 
even more challenging than fitting the fast-rotating stars. From this exercise it 
became clear that the line list used to fit the main sequence and turn-off 
stars is not fully adequate for red giants. Problems with the wavelengths 
and depths of the lines become clear. More work on the 
transitions affecting the near UV region is required. 

\begin{figure}
\begin{centering}
\includegraphics[width=7cm]{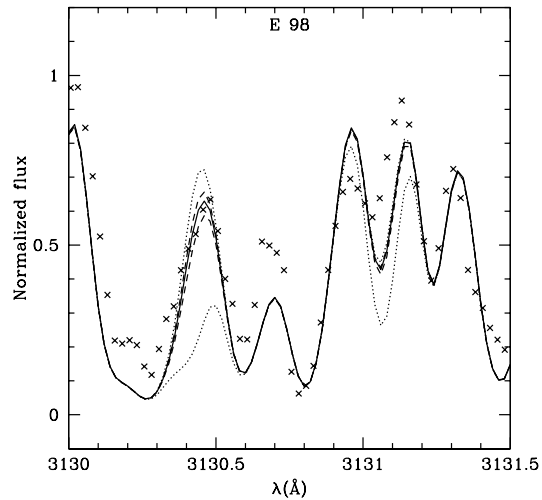}
\caption{Fitting the Be lines in the red giant E98. The crosses represent the 
observed spectrum, the solid line the best fit, the dashed lines represent 
variations of $\pm$ 0.20 dex from the best fit, and the dotted lines represent 
a spectrum without Be and one with A(Be) = 1.21.}
\label{fig:e98}
\end{centering}
\end{figure}

In particular, the \ion{Fe}{i} line adopted as the blend affecting the 
3131 \AA\@ Be line appears to be too strong here and with a wrong 
wavelength when compared to the observed spectra (see Fig. \ref{fig:e98}). 
For the dwarf stars, as discussed before, this blend has no significant effect. 
For metal-poor stars, such as the subgiants analyzed by \citet{GPP06}, 
the blending feature is also not significant. For our 
metal-rich cool giants, however, it prevents reliable abundances from being derived 
from this Be line. In this case, we relied solely on the 3130.4 \AA\@ feature.

The 3130.4 \AA\@ line, however, is very weak (indicating that Be is strongly 
depleted or even absent in giants) and strongly blended. In Fig.\@ 
\ref{fig:e98} an example of fitting the Be region in a red giant 
is given. The Be line only affects the wing of a very strong and heavily 
blended feature. The best fit is found when the wing of this feature in the 
synthetic spectrum is brought to the level of the wing in the observed 
spectrum. The figure suggests that an abundance cannot be derived in this 
case with an uncertainty better than $\pm$0.20 dex. 

One should also note that the abundances in this case are strongly 
dependent on broadening, blending features (their inclusion and also 
on the right abundance for the given element), and on the 
determination of the continuum. Thus, one should be very careful 
when interpreting the abundances derived for the red giants. In 
this case we consider that only an upper limit on the Be abundance 
can be determined. In addition, the difficulties discussed above mean that  
one should be very careful not to push the comparison between dwarfs 
and giants too far. The Be upper limits for the red giants 
are listed in Table \ref{tab:par}. 

Of all these issues, the level of the continuum is of particular concern. 
Different solutions could bring the wing of the line up or down, 
changing the best fit. Finding the correct level 
of the continuum in the near UV region is at best a very difficult 
task. However, by adopting very similar solutions for all the stars 
that have very similar atmospheric parameters, we believe that 
at least a reliable differential scale can be obtained.

Regarding the abundances of other elements, it is expected that the first dredge-up would 
change C and N abundances of the stars on the red giant branch (RGB). This near UV region is affected by many 
CN, NH, and CH lines, so this change in abundances could be 
an important issue. We calculated a synthetic spectrum for star E60 by 
adopting typical post dredge-up values for these elements, [C/Fe] = $-$0.15 
and [N/Fe] = +0.40 \citep[see e.g.][]{Sm09b}. A comparison with the best fit 
obtained using unaltered abundances showed only very weak effects, well within 
the $\pm$0.20 uncertainty of the fit.

\begin{figure}
\begin{centering}
\includegraphics[width=7cm]{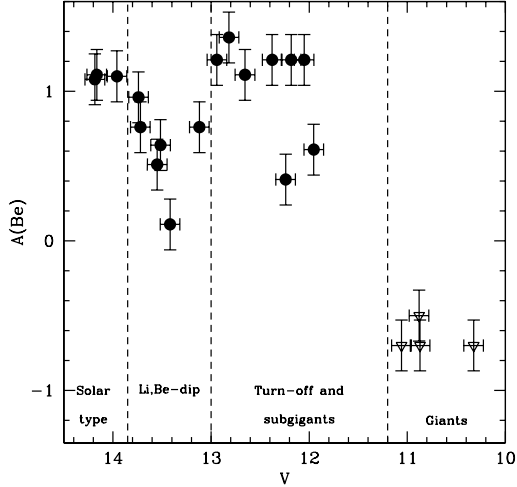}
\caption{The Be abundances of the sample stars, calculated 
in this work, as a function 
of the $V$ magnitude. Abundance determinations are shown 
as solid circles and upper limits as open triangles. The Be dip is clearly seen.}
\label{fig:bey}
\end{centering}
\end{figure}

\begin{figure}
\begin{centering}
\includegraphics[width=7cm]{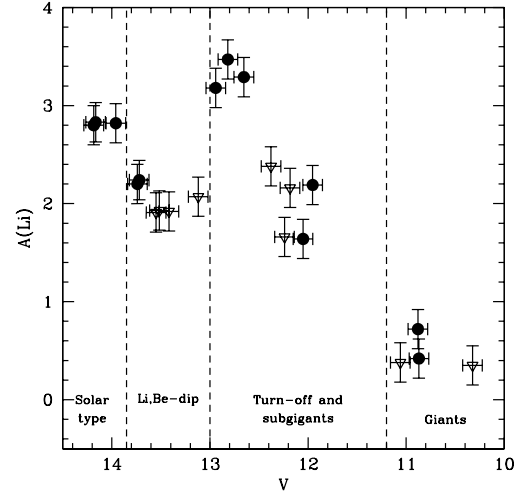}
\caption{The Li abundances of the sample stars, calculated in 
\citet{PRZ04},  as a function 
of the $V$ magnitude. Abundance determinations are shown 
as solid circles and upper limits as open triangles. }
\label{fig:liy}
\end{centering}
\end{figure}

The Be abundances of all sample stars are shown as a function of 
the $V$ magnitude in Fig. \ref{fig:bey}. For comparison, we also 
show a plot of the Li abundances determined by \citet{PRZ04} as 
a function of $V$ in Fig.\ \ref{fig:liy}. The evolutionary stages of the 
stars are indicated in the figures. It is clearly seen that the stars 
of the Li dip also define a clear Be dip.

\begin{table}
\caption{The uncertainties of the Be abundance introduced by the
  uncertainties of the atmospheric parameters and by uncertainty of the fitting 
itself related to the level of the continuum and the S/N.}
\centering 
\label{tab:sigma}
\begin{tabular}{ccccccc}
\noalign{\smallskip}
\hline\hline
\noalign{\smallskip}
Star & $\sigma_{\rm Teff}$ & $\sigma_{\rm log g}$ & $\sigma_{\rm \xi}$ &
 $\sigma_{\rm [Fe/H]}$ & $\sigma_{\rm fitt.}$  & $\sigma_{\rm total}$ \\
\hline
E 7 & $\pm$0.03 & $\pm$0.10 & $\pm$0.00 & $\pm$0.00 & $\pm$0.05 &  $\pm$0.12 \\
E 86 & $\pm$0.05 & $\pm$0.10 & $\pm$0.05 & $\pm$0.05 & $\pm$ 0.10 & $\pm$0.17 \\
E 60 & $\pm$0.15 & $\pm$0.05 & $\pm$0.03 & $\pm$0.05 & $\pm$ 0.20 & $\pm$0.26 \\
\noalign{\smallskip}
\hline
\end{tabular}
\end{table}

\subsection{Uncertainties and comparison with the literature}

The determination of Be abundances is affected by the uncertainties of 
the atmospheric parameters and by the uncertainty in the determination 
of the pseudo-continuum during the spectrum fitting. To conduct 
the calculations of the errors, we adopted 
the same uncertainties of the atmospheric parameters determined by 
\citet{PRZ04}, $\sigma_{\rm Teff}$ = $\pm$ 100 K, $\sigma_{\rm logg}$ = 
$\pm$ 0.30 dex, $\sigma_{\xi}$ = $\pm$ 0.30 km s$^{-1}$, and 
$\sigma_{\rm [Fe/H]}$ = $\pm$ 0.10 dex.

To estimate the effect of the atmospheric parameters, we can change 
each one by its own error, keeping the other ones with the original 
adopted values, and recalculate the abundances. Thus, we measure what 
effect the variation of one parameter has on the abundances. We assume 
the uncertainties introduced by each parameter to be independent of the 
other ones. The uncertainty due to the continuum was determined by 
estimating the sensitivity of the derived Be abundance on the adopted 
continuum level. This uncertainty is mostly related to the S/N of 
the spectrum. These effects are listed in Table \ref{tab:sigma}. The calculations 
were done for three representative stars, E7 as an example of slow-rotating  
star, E86 as an example of a fast-rotating star, and E60 as an example of 
a giant star. 

The Li and Be abundances derived for the star E5, which seem to be 
slightly above the solar values (see Table \ref{tab:par}), seem to suggest that the uncertainty 
on its temperature is slightly higher than for the other stars. For these 
hot stars, the effective temperature is also important for determining 
Be abundances. We thus believe that its Li and Be abundances are 
slightly overestimated and should be interpreted with care. 

As mentioned in \S \ref{sec:obs}, \citet{Ran02} has analyzed three 
main sequence stars of IC 4651 and determined their Be abundances. 
Two of their stars, E7 and E45, are also analyzed here. The 
atmospheric parameters derived by \citet{Ran02} are different from the ones 
adopted here. \citet{Ran02} calculated temperatures of 
6061 and 6016 K, while we adopted 6300 and 6350 K for stars E7 and E45, 
respectively. A gravity of log g = 4.44 was adopted by 
\citet{Ran02} for all stars they analyzed, while a value of 
log g = 4.30 was calculated for both E7 and E45 in \citet{PRZ04} and adopted here. In 
spite of these differences, the Be abundances of the two stars as 
derived by \citet{Ran02} agree with ours within the uncertainties; that is, \citet{Ran02} 
derived A(Be) = 1.11 and 1.16 for stars E7 and E45, while we 
obtained A(Be) = 1.11 and 1.08, respectively.

\section{Comparison with model predictions for rotating stars}\label{sec:dis}

\subsection{Input physics of the hydrodynamical stellar models}\label{sec:input}

\subsubsection{Hot side of the Li dip}\label{sec:hot}

In \citet{PRZ04} the Li abundance patterns in IC 4651 were compared with model 
predictions for rotating stars by \citet{CT99} and \citet{PTCF03}. These models 
include atomic diffusion, as well as the transport of angular momentum and 
chemicals by meridional circulation and shear turbulence following the formalism 
by \citet{Z92}, \citet{TZ97}, and \citet{MZ98} \citep[for more details and 
references see the original papers as well as][]{Decr09}. They are adequate for 
stars that lie on or originate in the hot side of the dip and inside which 
angular momentum transport by internal gravity waves emitted by convective envelopes 
is inefficient \citep{TC03}. These models were found to explain the 
corresponding Li data for both main sequence and giant stars in the intermediate-age 
IC 4651 \citep[see][]{PRZ04}, as they do for younger clusters like the Hyades, 
Coma Berenices, and Praesepe \citep[see e.g.][]{ipCT08}.

\begin{table*}
\caption{Rotation velocity, effective temperature, and surface abundances of Li 
and Be predicted for the stellar models that are still on the main sequence at 
the age of IC~4651 (models by Charbonnel \& Lagarde, in preparation).  See 
text for more details.}
\label{tab:models}
\centering
\begin{tabular}{c | c | c c | c c | c c | c c |}        
	\hline\hline                        
	M & $V_{ZAMS}$&\multicolumn{2}{c}{$V_{Age}$}& \multicolumn{2}{c}{T$_{\rm eff}$} & \multicolumn{2}{c}{A(Li)} &\multicolumn{2}{c}{A(Be)}   \\    
	$(M_{\odot})$ &$(km\ s^{-1})$ &\multicolumn{2}{c}{ $(km\ s^{-1})$} &\multicolumn{2}{c}{(K) } & & &  &\\	
	&&1.2 Gyr & 1.7 Gyr &1.2 Gyr & 1.7 Gyr &1.2 Gyr & 1.7 Gyr &1.2 Gyr & 1.7 Gyr \\
	\hline\hline                                   
	 1.3 & 50 & 19.4 & 16.9 & 6532 & 6507 & 2.05 & 1.51 & 1.02 & 0.77\\    
	       & 80 & 28.3 & 24.4 & 6535 & 6528 & 1.24 & 0.37& 0.62 & 0.22 \\
	       & 110 & 36.2 & 30.9 & 6520 & 6483 & 0.47 & -0.59 &0.2 & -0.29 \\
	\hline
       1.35 & 50 & 30.3 & 27.6 & 6687 & 6663 & 2.69 & 2.33 & 1.26 & 1.10 \\
                & 80 & 28.6 & 24.5 & 6683 & 6670 & 1.16 & 0.24 & 0.57& 0.15 \\
                & 110 & 36.6 & 31.5 & 6685 & 6676 & 0.43 & -0.70 & 0.20 & -0.35 \\
	\hline
       1.4 & 50 & 42.7 & 41.4 & 6834 & 6762 & 3.24 & 3.22 & 1.41 & 1.41 \\
       	    & 80 & 67.8 & 67.8 & 6824 & 6764 & 3.20 & 3.10 & 1.41 & 1.39 \\
	    & 110 & 96 & 95.2 & 6826 & 6781 & 3.12 & 2.94 & 1.39 & 1.33 \\
	\hline
       1.5 & 50 & 40.5 & 37.5 & 7094 & 6857 & 3.23 & 3.18 & 1.41 & 1.40 \\
              & 80 & 66.5 & 62.7 & 7174 & 6948 & 3.13 & 2.94 & 1.40 & 1.34\\
              & 110 & 92.2 & 87.8 & 7136 & 6900 & 2.97 & 2.66 & 1.35 & 1.21\\              
	\hline 
	\end {tabular}
\end{table*}

Meanwhile new stellar models within the same mass range (i.e., for stars more 
massive than 1.2~M$_{\odot}$) were computed for other purposes with the 
stellar evolution code STAREVOL V3.1 (Charbonnel \& Lagarde, in preparation) 
using the same prescriptions for the transport of angular momentum and chemicals 
by meridional circulation and shear turbulence as in the above-mentioned papers. 
Atomic diffusion is also included in the form of gravitational settling, as well as 
that related to thermal gradients, using the formulation by \citet{Paquette86}.
As far as the input physics is concerned, the differences with the models by 
\citet{CT99} and \citet{PTCF03} are the following: (1) the reference solar composition 
is the one derived by \citet{AspGS06}, except for Ne for which the 
recommendations by \citet{Cunha06} are followed; (2) new OPAL opacity tables are used according 
to the chemical mixture; (3) thermohaline mixing is accounted for following 
\citet{CZ07}, a process that only becomes efficient when the stars reach the 
so-called luminosity bump on the RGB (see \S \ref{sec:giant}). 

Results of these recent computations for initial stellar masses 
of 1.3, 1.35, 1.4, 1.5, 1.8, and 2~M$_{\odot}$ for Z = 0.014 are presented. In view of the remaining 
uncertainty on the age of IC~4651, we provide in Table \ref{tab:models} relevant 
model characteristics at 1.2 and 1.7~Gyr for the stars that are still on the 
main sequence at that time. In all cases various initial rotation velocities on 
the zero-age main sequence were considered. (The 1.3, 1.35, 1.4, and 1.5~M$_{\odot}$ models are computed for 
V$_{ZAMS}$ = 50, 80, and 110 km s$^{-1}$; the 1.8~M$_{\odot}$ models are computed for V$_{ZAMS}$ 
= 80, 110, and 180 km s$^{-1}$, and the 2~M$_{\odot}$ models for V$_{ZAMS}$ = 110, 180, and 250 km 
s$^{-1}$.) The efficiency of magnetic braking  applied to stars arriving on the main sequence is determined 
according to the work of \citet{Gaige93} following the braking 
law by \citet{Kawaler88} \citep[for more details see][]{PTCF03}. In practice, magnetic 
braking only has to be applied to the 1.3, 1.35, and 1.4~M$_{\odot}$ stars that have 
thicker convective envelopes than more massive stars. As in previous works, the aim was to 
reproduce the rotation velocity distribution observed at the age of the Hyades. 
Theoretical rotation velocities at the age of IC~4651 are given in Table \ref{tab:models} and span the 
velocity range of our sample stars. In all cases the surface rotation velocity 
also changes because of secular evolution, and it is reduced to only a few km s$^{-1}$ 
when the models are on the subgiant branch due to the radius expansion.
  
\subsubsection{Cool side of the Li dip}\label{sec:cool}

For less massive main-sequence stars lying on the cool side of the Li dip that 
have deeper convective envelopes, internal gravity waves are efficiently generated 
and participate in the extraction of internal angular momentum, along with 
meridional circulation and shear turbulence \citep{TC03}. In this case, the 
appropriate models are those by \citet{TC05} and \citet{ChT05}, which take all these effects 
into account. Unfortunately the 1.0~M$_{\odot}$ models of \citet{ChT05} have an 
effective temperature at the age of the IC~4651 (i.e., 5597~K), lower than our cooler 
main sequence stars, and they cannot be used for comparison with the present sample. Obviously 
abundance determinations for such relatively cool objects would be extremely valuable for further tests of 
the models. Only the 1.2~M$_{\odot}$ model computed by \citet{TC05} is 
available for comparison purposes (see \S \ref{sec:main}). 
It has an initial rotation velocity on the zero age main sequence of 50~km s$^{-1}$, 
and surface spin-down follows Kawaler (1988) as explained in \S~\ref{sec:hot}.

\subsubsection{Initial values for Li and Be}\label{sec:initial}

In the following we compare the predictions of the hydrodynamical models described above 
with the observations of both Li and Be in IC~4651. 
Results are presented separately for the main sequence and turnoff 
stars (\S \ref{sec:main}) and for more evolved stars (\S \ref{sec:giant}). In the 
following figures, we use the $V$ magnitudes to divide the stars according to their 
evolutionary stages; stars fainter than V = 12.6 are at the main sequence, stars 
between 12.6 and 11.8 are at the upper turn-off or at the subgiant branch, and 
stars brighter than 11.8 are giants.

For the initial values of A(Li) and A(Be) used in our computations we chose the 
meteoritic abundances, i.e., 3.25 \citep{AspGS06} and 1.41 \citep{Lod03}, respectively. 
This Li value is actually lower than the highest Li abundance determined among our sample stars, 
which induces an artificial shift within the comparison between predictions and data. 
For this reason we increase, within the figures, the theoretical Li points by 0.16~dex. 
This corresponds to assuming that the highest Li abundance derived among our sample stars (E5, 
A(Li)=3.47~dex) is the initial abundance of the sample stars.

\subsection{Comparisons with observed features on the main sequence}\label{sec:main}

\begin{figure}
\centering
\includegraphics[width=7cm]{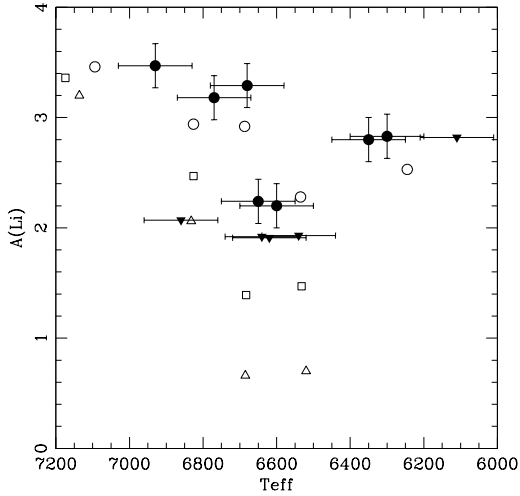}
\caption{Li abundances in IC~4651 main sequence and turnoff stars fainter than V $\sim$12.6 
(black points and triangles for actual determinations and upper limits, respectively). Open 
circles, squares, and triangles show model predictions for initial rotation velocities
of 50, 80, and 110~km s$^{-1}$, respectively. For comparison purposes the theoretical Li 
points are increased by 0.16~dex to account for the difference between the meteoritic value and the 
highest Li abundance determined in our sample stars. See text for more details.
}
\label{fig:Li_Teff_1p2Gyr}
\end{figure}

\begin{figure}
\centering
\includegraphics[width=7cm]{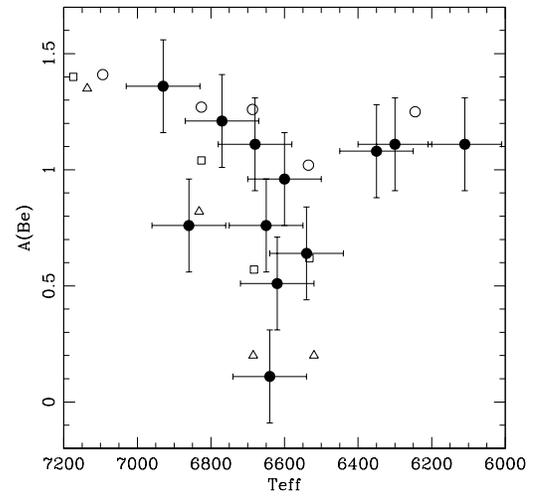}
\caption{Same as Fig.\ref{fig:Li_Teff_1p2Gyr} for observed and theoretical Be abundances.}
\label{fig:Be_Teff_1p2Gyr}
\end{figure}

Figures \ref{fig:Li_Teff_1p2Gyr} and \ref{fig:Be_Teff_1p2Gyr} compare the predictions 
of the stellar models described in \S \ref{sec:input} to the Li and Be 
observed abundances of the main-sequence and turn-off stars of our IC~4651 sample as a function of the 
effective temperature. The models provide an estimate 
of the expected dispersion in surface abundances due to a dispersion in initial 
rotation velocities. On the cool side of the dip, only one theoretical point is 
available in the temperature range considered in the present study 
\citep[1.2~M$_{\odot}$, V$_{ZAMS}$ = 50~km s$^{-1}$ model of][]{TC05}.

\begin{figure}
\centering
\includegraphics[width=7cm]{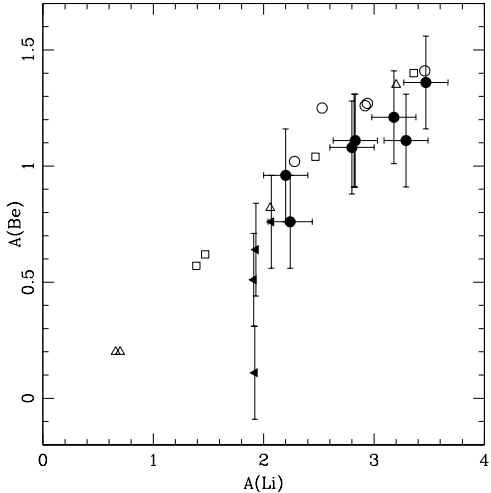}
\caption{Li and Be abundances in IC~4651 main sequence and turnoff stars fainter than V $\sim$12.6 
(black points and triangles for Li actual determinations and upper limits, respectively). Open 
circles, squares, and triangles show model predictions for initial rotation velocities
of 50, 80, and 110~km s$^{-1}$, respectively. For comparison purposes the theoretical Li 
points are increased by 0.16~dex to account for the difference between the meteoritic value and the 
highest Li abundance determined in our sample stars. See text for more details.
}
\label{fig:Be_vs_Li_1p2Gyr_Liini3p47}
\end{figure}

It is possible to note on these graphs (see also Fig.~\ref{fig:Be_vs_Li_1p2Gyr_Liini3p47}) 
that the observational behavior of the Be abundances 
follows that of the Li abundances closely. The Li dip is also a Be dip, 
confirming previous literature results \citep{BK02,BAK04}.
In a narrow range in effective temperature centered on $\sim$ 6650~K, stars present 
both an Li and a Be depletion. Important is that Li, which is more fragile (it burns indeed 
through proton-captures at $\sim$ 2.5$\times 10^{6}$ K compared to $\sim$ 3.5$\times 10^{6}$ K for Be), 
is more strongly depleted within the dip. In most of the stars lying in this region, 
only upper limits could be derived for Li, while Be is always detected.

The predictions of the models described in \S~\ref{sec:input} successfully account 
for the Li patterns as a function of effective temperature, namely, 
the presence of a large fraction of stars with Li upper 
limits between $\sim$ 6500 and 6800~K (i.e., within the dip) and the higher Li 
values on both the cool and hot sides of the dip (see Fig.~\ref{fig:Li_Teff_1p2Gyr}). 
They also reproduce the Be behavior remarkably well in the whole effective temperature range considered 
(see Fig.~\ref{fig:Be_Teff_1p2Gyr}). The success of the models is confirmed when plotting
Be versus Li as shown in Fig.~\ref{fig:Be_vs_Li_1p2Gyr_Liini3p47}. As noted in \S~4.1.3, the theoretical Li 
values have been shifted by 0.16~dex to account for the difference between the meteoritic Li and the highest 
Li abundance determined in our sample stars; however, what really matters for constraining the models is the slope of 
the trend between Li and Be. As can be seen in Fig.~\ref{fig:Be_vs_Li_1p2Gyr_Liini3p47}, 
the theoretical and observational slopes are found to agree very nicely. Linear fits of 
the observational values, accounting for errors in both axes, have slopes of +0.58 $\pm$ 0.09, 
if all detections and upper limits are considered, of +0.45 $\pm$ 0.05 
if we only exclude star E34 (the one with the smaller Be abundance), and of +0.35 $\pm$ 0.08 if 
we consider only the detections.  For the theoretical values we calculate a slope of +0.45 
considering all points. If we restrict the theoretical values to the ones with A(Li) $>$ 2.0, which is 
the range where the observational detections are, a slope of +0.33 is obtained. These 
values agree very well with the observed ones. Similar correlated 
depletions of Li and Be have also been observed in field and cluster stars by \citet{DBS98}, \citet{BAKDS04}, 
and \citet{Ran07}.

On the cool side of the dip (i.e., below $\sim$ 6400~K), we find no Be dispersion 
thus confirming previous results by \citet{BAKDS04} and \citet{Ran02,Ran07}. This 
provides further constraints to models of rotating low-mass (i.e., below $\sim$ 1.25~M$_{\odot}$), 
stars including the transport of angular momentum by gravity waves that should assume 
a range in initial rotation 
velocities (Charbonnel et al. in preparation)\footnote{\citet{Ran02} compared 
the Li and Be abundances determined in early G-type stars in old clusters, among 
which IC~4651 with predictions of models including gravity waves by 
\citet{MoSc00}. In these models, rotation-induced processes are ignored, and 
waves are invoked as the only source of mixing for chemicals in the stellar 
radiative zone. This is very different from the view proposed by Charbonnel 
\& Talon. In this case indeed gravity waves transport angular momentum, together 
with meridional circulation and turbulence, but the transport of chemicals is 
essentially caused by rotation-induced mixing. See the original papers for more 
detailed information.}. However, we can already note that models of solar-mass stars computed with 
a large range of initial angular momentum (V$_{ini}$ from 15 to 110~km s$^{-1}$) 
and including the effects of internal gravity waves predict very small Li dispersion 
at the age of IC~4651 \citep[see][]{ChT05}, and thus even more modest Be dispersion.
This is caused by internal gravity waves that dominate the transport of angular momentum 
in stars lying on the red side of the dip and enforce quasi solid-body rotation within the stellar interior. As a result, the surface 
Li and Be depletions are expected to almost be independent of the initial 
angular momentum distribution (i.e., of the initial rotation velocity), implying very low
surface abundance variations from star to star. 
On the other hand, in more massive (i.e., higher than $\sim$ 1.25~M$_{\odot}$) stars
the efficiency of internal gravity waves dramatically drops and internal differential rotation is 
expected to be maintained along the stellar life under the effects of meridional circulation and turbulence.
Consistently, variations in the initial angular momentum from star to star lead 
to more Li and Be dispersion, as required by the observations.

\subsection{Comparisons with features observed in subgiant and giant stars}\label{sec:giant}

\begin{figure}
\centering
\includegraphics[width=7cm]{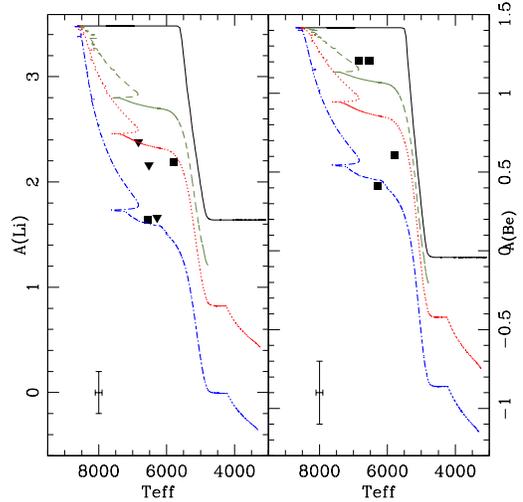}
\caption{Li and Be abundances in IC~4651 subgiant stars with T$_{\rm eff}$ between $\sim$ 
6830 and 5800~K and V between 12.6 and 11.8. Theoretical predictions for the 
surface Li and Be evolution are shown for the 1.8~M$_{\odot}$ star (see text 
for more details). The solid line is for the classical case, while the other 
lines present the results for the rotating models computed with different 
initial rotation velocities (80, 110, and 180 km s$^{-1}$, respectively, 
for dashed, dotted, and dot-dashed lines). The Li tracks are shifted by 0.16~dex as explained in the text.}
\label{fig:subgiantstars}
\end{figure}

\begin{figure}
\centering
\includegraphics[width=7cm]{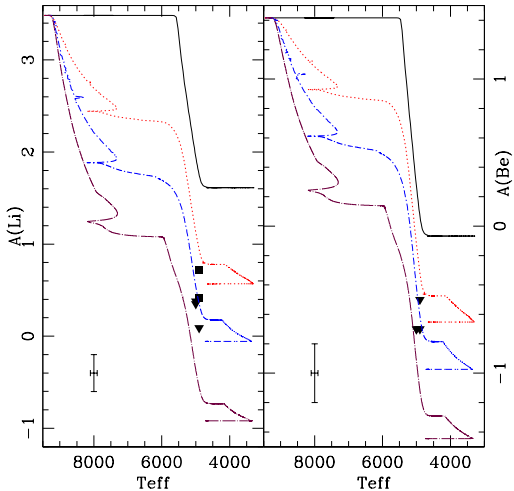}
\caption{Li and Be abundances in IC~4651 evolved stars brighter than V $\sim$ 12.6. 
Predictions are shown for the 2~M$_{\odot}$ model. 
The solid line is for the classical case, while the other lines present the results for the 
rotating models computed with different initial rotation velocities (110, 180, and 
250 km s$^{-1}$ for dotted, and dot-dashed, and dot-long dashed lines, respectively). 
The Li tracks are shifted by 0.16~dex as explained in the text.}
\label{fig:giantstars}
\end{figure}

Let us now focus on the post-main-sequence evolution. Figures~\ref{fig:subgiantstars} 
and \ref{fig:giantstars} compare the Li and Be abundances of our subgiant and giant 
sample stars, as a function of the effective temperature (which now depicts evolutionary stage), 
with predictions of our most massive models computed by assuming different 
initial rotation velocities. Standard (i.e., non-rotating) predictions are also shown. 
The subgiant stars (with T$_{\rm eff}$ between $\sim$ 6830 and 5800~K) are 
compared with the 1.8~M$_{\odot}$ model predictions, while the relevant tracks for
the giants (with T$_{\rm eff}$ at $\sim$ 4900-5000~K) are the 2M$_{\odot}$ ones.

In the classical (non-rotating) case, the surface depletion of both Li and Be begins 
relatively late in T$_{\rm eff}$ within the Hertzsprung gap (i.e., at $\sim$ 5500~K). This 
depicts dilution of the external layers with nuclear-processed material when the 
convective stellar envelope deepens in mass. However and as shown in Fig.~\ref{fig:subgiantstars}, 
actual dilution starts earlier than predicted classically, 
as already revealed by Li data both in open clusters like IC~4651 \citep{PRZ04} and NGC~3680 
\citep{PRP01}, but also in field stars \citep{Alschuler75,Brown89,Leb99,DoN00,PTCF03}. 
As can be seen in Fig.~\ref{fig:subgiantstars}, the Be depletion also occurs earlier than 
predicted in the classical case. This simply means that the Li- and Be-free regions 
are larger in real main sequence stars than classically predicted.

This favors the complete hydrodynamical models, since rotation-induced mixing 
acting on the main sequence enlarges the Li- and Be-free regions inside the stars, implying earlier (i.e., at 
higher T$_{\rm eff}$ inside the Hertzsprung gap) Li and Be depletion at the surface of subgiant stars,  
as well as lower Li and Be values after dredge-up in giant stars 
\citep[see also][ for model comparison with Li data for field stars]{PTCF03}.
The behavior of both elements predicted in the rotating case nicely matches the data 
for the subgiant and giant stars in IC~4651, and a dispersion in initial rotation 
velocity explains the observed abundance dispersion along the evolutionary 
sequence well.

Lithium abundances of the red giants of IC 4651 are discussed 
in detail by \citet{PRZ04}. As can be seen from Table \ref{tab:par}, Li 
was detected in only two giants, although all five of them have very similar atmospheric parameters. 
To explain this, \citet{PRZ04} suggests that the Li observed 
in these stars might have been produced in their interior after the 
first dredge-up. We can see, however, that this is not necessary, since in the corresponding stellar mass domain, 
differences in initial rotation velocities are expected to lead naturally to a significant dispersion 
($\sim$ 1.6 dex for the range in initial rotation assumed here) in post-dredge-up Li values, 
accounting nicely for the observed data.

\citet{PRZ04} also discusses the issue of the evolutionary status of the giant stars in IC~4651.
Their proper classification is based only on their position at the CMD and is not straightforward. 
These stars could indeed be either on the first ascent RGB or at the clump. 
As mentioned in \S \ref{sec:hot}, the new models presented here are computed to include 
thermohaline mixing following \citet{CZ07}. This process only becomes efficient 
when the stars reach the bump on the RGB. It then leads to an additional decrease in both the Li and Be surface abundance 
at T$_{\rm eff}$ $\sim$ 4250~K, as can be seen in Fig.~\ref{fig:giantstars}. Clump stars 
are thus expected to have lower Li and Be surface abundances than their RGB counterparts.
However, for a 2~M$_{\odot}$ star the effect is relatively modest (see also Charbonnel 
\& Lagarde, in preparation), and the theoretical predictions cannot help us disentangling  
the evolutionary status of the giants.


\section{Conclusions}\label{sec:con}

Beryllium abundances were calculated for twenty-one main sequence, turn-off, 
subgiant, and giant stars belonging to the open cluster IC 4651. This is the first time that 
such an analysis has been carried out along the whole evolutionary sequence of an open cluster. 

The Be abundances are found to closely follow the behavior of 
the Li abundances as determined by \citet{PRZ04}. The coolest main sequence 
stars observed present no Be abundance dispersion, 
and a well-defined Be dip overlaps the well-documented Li dip for F-type stars.
The Be abundances of post-main-sequence stars confirm 
that first-dredge up dilution starts earlier than the expected classically and present a significant dispersion.
This is consistent with the Li behavior in evolved stars in IC~4651, in NGC~3680, and in the field.

Our observations are compared with theoretical predictions. 
For the main sequence and turnoff stars lying on the hot side of the Li-dip, as well as for the evolved 
stars, we compared the observations with results of new hydrodynamical 
stellar models calculated with STAREVOL V3.1 (Charbonnel \& Lagarde, in preparation). 
These models include the effects of atomic diffusion, meridional circulation, shear turbulence, and 
thermohaline mixing. (Wave-induced transport is negligible in the case of these stars that 
have very narrow convective envelope while on the main sequence.) 
For the main sequence stars lying on the cool side of the Li dip, we compared the observations 
with the 1.2 M$_{\odot}$ model of \citet{ChT05} which also includes the transport of 
angular momentum by internal gravity waves.

The models reproduce all the Li and Be features very nicely along the CMD of IC~4651. 
The dispersion of Li and Be abundances on the blue side of the dip and in evolved stars is very well 
explained when accounting for a dispersion in the initial values of the stellar rotational velocity 
as observed in young open clusters.
On the other hand, the lack of abundance dispersion in lower mass stars is expected to be caused by 
the impact of internal gravity waves.

The success in explaining the Li and Be abundances along the whole 
evolutionary sequence of IC~4651 is very encouraging. It shows that important steps 
have been taken towards the proper understanding of the physical mechanisms 
acting during the stellar evolution.

\begin{acknowledgements}

This research has made use of the WEBDA database, operated at the
Institute for Astronomy of the University of Vienna, of NASA's 
Astrophysics Data System, and of the Simbad
database operated at CDS, Strasbourg, France. R.S. acknowledges 
FAPESP PhD (04/13667-4) and post-doc fellowships (08/55923-8). C.C. and N.L. acknowledge financial 
support from the FNS (Switzerland) and from the Programme National de Physique 
Stellaire of CNRS/INSU (France).

\end{acknowledgements}

\bibliographystyle{aa}
\bibliography{../../Tese/rsmiljanic}


\end{document}